\documentclass[a4paper,11pt]{article}
\usepackage{pos}
\usepackage{graphicx}
\usepackage{amsfonts, amsmath, amsthm}
\usepackage{mathtools}
\usepackage{caption}
\usepackage{subcaption}
\usepackage{xcolor}

\newcommand{\cbo}{\chi_{b1}}

\title{NRQCD Bottomonium at non-zero temperature using time-derivative moments}

\author*[a]{Rachel Horohan D'arcy}
\author[c]{Gert Aarts}
\author[c]{Chris Allton}
\author[b]{Ryan Bignell}
\author[c]{Timothy Burns}
\author[d]{Benjamin Jaeger}
\author[e]{Seyong Kim}
\author[f]{Maria Paola Lombardo} 
\author[b]{Sinéad Ryan}       
\author[c]{Antonio Smecca}
\author[a,b]{Jon-Ivar Skullerud}

\affiliation[a]{Department of Theoretical Physics, National University of Ireland Maynooth, Maynooth, Co Kildare, Ireland}
\affiliation[b]{School of Mathematics and Hamilton Mathematics Institute, Trinity College Dublin, Ireland}
\affiliation[c]{Department of Physics, Swansea University, Swansea, SA2 8PP, United Kingdom}
\affiliation[d]{CP3-Origins and Danish IAS, Department of Mathematics and Computer Science, University of Southern Denmark, Campusvej 55, 5230 Odense M, Denmark}
\affiliation[e]{Department of Physics, Sejong University, Seoul 05006, Korea}
\affiliation[f]{INFN, Sezione di Firenze, 50019 Sesto Fiorentino (FI), Italy}

\emailAdd{rachel.horohandarcy.2018@mumail.ie}
\emailAdd{g.aarts@swansea.ac.uk}
\emailAdd{c.allton@swansea.ac.uk}
\emailAdd{bignellr@tcd.ie}
\emailAdd{t.burns@swansea.ac.uk}
\emailAdd{jaeger@imada.sdu.dk}
\emailAdd{skim@sejong.ac.kr}
\emailAdd{mariapaola.lombardo@inf.infn.it}
\emailAdd{ryan@maths.tcd.ie}
\emailAdd{antonio.smecca@swansea.ac.uk}
\emailAdd{jonivar@thphys.nuim.ie}

\abstract{A well-known challenge for the lattice community is calculating the spectral function from the Euclidean correlator. We have approximated the spectral function and derived the mass and thermal width of particles through the time derivatives of the lattice correlator moments.
We have focused on extracting the properties of bottomonium states, specifically $\Upsilon$ and $\chi_{b_1}$. We will give an overview of the time-derivative moments approach and present results for the temperature dependence of the mass and width of both bottomonium states. The zero temperature results are consistent with experimental values, while results at higher temperatures are similar to those obtained using other methods.}

\FullConference{The 41st International Symposium on Lattice Field Theory (LATTICE2024)\\
 28 July - 3 August 2024\\
Liverpool, UK\\}

\begin{document}
\maketitle

\section{Introduction}

Heavy quarkonium states are important probes for quark-gluon plasma (QGP) and can be used as a thermometer for relativistic heavy-ion collisions \cite{Brambilla:2010cs}. It is therefore important to study bottomonium states above the deconfining temperature.
The b-quark has a mass that is larger than other energy scales and can be approximated as a non-relativistic particle. Non-relativistic QCD (NRQCD) is an effective field theory which approximates fully relativistic QCD by expanding the Lagrangian in powers of the heavy quark velocity in the bottomonium rest frame, $v = |\textbf{p}|/m_b$ \cite{Lepage:1992tx, Brambilla:2004jw}. The quark and anti-quark decouple and their propagators are then obtained as solutions to an initial value problem. This results in a Euclidean correlator, 
\begin{equation}
    G(\tau;T) = \int_{\omega_{min}}^{\infty} \frac{\text{d}\omega}{2\pi} K(\tau,\omega) \rho(\omega;T) \,,
    \label{eq:nrmesoncorr}
\end{equation}
with the NRQCD kernel is $K(\tau,\omega) = e^{-\omega\tau}$, and $T=(a_\tau N_\tau)^{-1}$. The challenge we now face is the ill-posed problem of reconstructing the spectral function, $\rho(\omega)$, from the discrete lattice correlator $G(\tau)$. 
We present a new method to study the spectra of heavy quarkonia. We analyse the correlator without attempting to reconstruct the full spectral function by calculating the central moments of the correlator after assuming the shape of the spectral function to be a sum of Gaussians. We use this method to study the bottomonium spectrum at finite temperature.  
\section{Lattice Setup}
We use FASTSUM's "Generation 2L" ensembles with anisotropic lattice spacings, using a Symanzik-improved anisotropic gauge action and an improved Wilson fermion action. Parameter details are in Table \ref{tab:latticeparams}. The temperature $T=(a_\tau N_\tau)^{-1}$ is varied by changing the number of temporal lattice sites $N_\tau$ see Table \ref{tab:Gen2l_temps}, and the pseudocritical temperature is $T_c=167$MeV. For details of the ensembles see Refs \cite{Aarts:2020vyb, Aarts:2022krz}.
\begin{table}
    \centering
    \begin{tabular}{c|cccccccc}
          Gen  & $N_f$ & $\xi$ & $a_s$ (fm) & $a_\tau^{-1}$(Gev) & $m_\pi$(MeV) & $N_s$ & $L_s$(fm) & $T_c(MeV)$\\ \hline
            2L & $2+1$  & 3.45 & 0.112& 6.08 & 240 & 32 & 3.58 & 167 \\ 
        \end{tabular}
        \caption{Lattice parameters for FASTSUM ensembles Generation 2L: Number of flavours $N_f$, spatial lattice spacing $a_s$, inverse temporal lattice spacing $a_\tau^{-1}$, anisotrophy $\xi=a_s / a_\tau$, pion mass $m_\pi$, number of lattice sites in the spatial direction $N_s$, spatial extent $L_s$, and the pseudocritical temperature $T_c$.}
    \label{tab:latticeparams} 
\end{table}
\begin{table}
    \centering
        \begin{tabular}{c|ccccccccccc}
            $N_\tau$ & 128 & 64 & 56 & 48 & 40 & 36 & 32 & 28 & 24 & 20 & 16 \\ \hline
             T(MeV) & 47 & 95 & 109 & 127 & 152 & 169 & 190 & 218 & 253 & 304 & 380 
        \end{tabular}
    \caption{Temperature corresponding to each $N_\tau$ for the Gen2L ensemble.}
    \label{tab:Gen2l_temps}
\end{table}
\section{Time-derivative Moments}
There are a number of methods that have been devised to reconstruct the spectral function $\rho(\omega;T)$, from the correlator in Equation \eqref{eq:nrmesoncorr} \cite{Spriggs:2021dsb}. Each approach comes with its own strengths and limitations. For a review and comparison of these methods see the upcoming paper \cite{Review}. Here we present a new method to extract the mass and thermal width without reconstruction of the spectral function. We approximate the spectral function, $\rho(\omega;T)$ by a finite sum of Gaussian functions, \cite{Spriggs:2021jsh, Larsen:2019bwy}, 
\begin{equation}
      \rho(\omega;T) \propto \sum_{i=0}^{N} e^{-\frac{(\omega-m_i)^2}{2\Gamma_i^2}},
      \label{eq:sumgaussspectfunc}
\end{equation}
and use the first and second moments of the resulting correlator to calculate the mass, $M$, and thermal width, $\Gamma$, of the bottomonium. Here we assume that the Gaussian associated with each state is centered at the mass, $M_i$, of that state, and the thermal width, $2\sqrt{2\text{log}(2)}\Gamma_i = \text{FWHM}_i$ of each Gaussian.\\

\paragraph{Mass:} The calculation of the mass, $M$, is most clearly illustrated in the limit $\Gamma^2 \to 0$ where the spectral function becomes a sum of delta fucntions, 
\begin{equation}
    \rho(\omega;T) \propto \sum_{i=0}^{N} \delta(\omega-m_i).
    \label{eq:sumdeltaspecfunc}
\end{equation}
Substituting \eqref{eq:sumdeltaspecfunc} into \eqref{eq:nrmesoncorr} with the NRQCD kernel gives a correlator of the form:
\begin{equation}
    G(\tau;T) \propto \sum_{i=0}^N \int \frac{\text{d}\omega}{2\pi} e^{-\omega\tau} \delta(\omega-m_i) = \frac{1}{2\pi} \sum_{i=0}^N e^{-m_i\tau}.
    \label{eq:sumdeltacorr}
\end{equation}
Now we consider the first central moment of the correlator and in the ground state limit. The mass $M$ of this state can be determined using
\begin{align}
    m = M_D(\tau) =& - \frac{G'(\tau;T)}{G(\tau;T)} = - \frac{\int -\frac{\text{d}\omega}{2\pi}\omega e^{-\omega\tau} \delta_{\omega,m}}{\int \frac{\text{d}\omega}{2\pi} e^{-\omega\tau} \delta_{\omega,m}},
    \label{eq:GprimeoverG} \\
    \intertext{or equivalently,}
    m = M_L(\tau) =& \frac{\partial(\text{log}(G(\tau;T)))}{\partial\tau}.
    \label{eq:derlogG}
\end{align}
Equations \eqref{eq:GprimeoverG} and \eqref{eq:derlogG} are identical in the continuum but give different results for discrete derivatives, and we have found \eqref{eq:derlogG} gives more reliable results. We use the difference between $M_L$ and $M_D$ to estimate systematic errors.
\\

\paragraph{Width:} For the thermal width we substitute the Gaussian spectral function \eqref{eq:sumgaussspectfunc} into Equation \eqref{eq:nrmesoncorr} with the NRQCD kernel, giving us a correlator,
\begin{equation}
    G(\tau;T) \propto \sum_{i=0}^{N} \int \frac{\text{d}\omega}{2\pi} e^{-\omega\tau}e^{-\frac{(\omega-m)^2}{2\Gamma^2}}.
\end{equation}
The spectral width $\Gamma^2$ can be equated to the variance of the Gaussian variable $\omega$, $\Gamma^2 = \operatorname{Var}\left< \omega \right>$. We take a weighted average of $\omega$ and $\omega^2$, Equation \eqref{eq:Gammasquared_weightedaverage}, where $e^{-\omega\tau}e^{-\frac{(\omega-m)^2}{2\Gamma^2}}$ is thought of as a "weighted spectral function".
\begin{align}
    \Gamma^2 =& \left<\omega^2\right> - \left<\omega\right>^2\\
     = & \frac{\int \frac{\text{d}\omega}{2\pi} \omega^2 e^{-\omega\tau}e^{-\frac{(\omega-m)^2}{2\Gamma^2}}}{\int \frac{\text{d}\omega}{2\pi}e^{-\omega\tau}e^{-\frac{(\omega-m)^2}{2\Gamma^2}}} - \left(\frac{\int \frac{\text{d}\omega}{2\pi} \omega e^{-\omega\tau}e^{-\frac{(\omega-m)^2}{2\Gamma^2}}}{\int \frac{\text{d}\omega}{2\pi}e^{-\omega\tau}e^{-\frac{(\omega-m)^2}{2\Gamma^2}}} \right)^2 ,\\
    \intertext{and we can see that this is equal to}
    \Gamma^2 = \Gamma^2_D(\tau) = &\frac{G''(\tau;T)}{G(\tau;T)} - \left(\frac{G'(\tau;T)}{G(\tau;T)}\right)^2 \label{eq:Gammasquared_weightedaverage} ,\\
    \intertext{or equivalently,}
    \Gamma^2 = \Gamma^2_L(\tau) = & \frac{\partial^2(\text{log}(G(\tau;T)))}{\partial\tau^2} . \label{eq:secondderlogG}
\end{align}
Similarly as for the mass, Equations \eqref{eq:Gammasquared_weightedaverage} and \eqref{eq:secondderlogG} are identical in the continuum, but give different results for discrete derivatives, and we have found that \eqref{eq:secondderlogG} gives more reliable results. We use the difference between $\Gamma_L$ and $\Gamma_D$ as an estimate of systematic errors.
\\

\paragraph{Fits:} We now go beyond the single state approximation and fit each of $M_L(\tau)$ and $\Gamma^2_L(\tau)$ to an appropriate function to get a value for $M_L$ and $\Gamma^2_L$ at each temperature. A spectral function, $\rho(\omega)$, as a sum of Gaussians leads to a correlator, $G(\omega)$, which is a sum of exponential functions. Each Gaussian and therefore each exponential in the sum corresponds to distinct bound states.   
\begin{align*}
    \rho(\omega) \propto \sum_{i=0}^{N} A_i e^{-\frac{(\omega-m_i)^2}{2\Gamma_i^2}} \hspace{1em }\Rightarrow{}  \hspace{1em} G(\tau) = \sum_{i=0}^{N} A_i e^{-m_i\tau+\Gamma_i^2\tau^2/2} .\\
\end{align*}
We focus on the ground state and assume it is well separated from the excited states and hence at large $\tau$, $\exp(-\Delta m_i \tau + \Delta\Gamma_i^2\tau^2/2)$ becomes small, with $\Delta m_i = m_i - m_0$ and $\Delta\Gamma_i^2 = \Gamma^2_i - \Gamma^2_0 $. We also assume $\Delta\Gamma^2\tau^2 < \Delta M \tau$ on our finite lattice. Further we don't see evidence of the $\Delta\Gamma^2_i$ term dominating in the data. With this assumption we are able to factorise the ground state out to get
\begin{align}
    G(\tau) = &e^{-m_0\tau+\Gamma_0^2\tau^2/2}\left(1+\sum_{i=1}^{N}\frac{A_i}{A_0}e^{-\Delta m_i\tau+\Delta\Gamma_i^2\tau^2/2}\right) ,\\
    \text{log}(G(\tau)) 
     = &-m_0\tau+\Gamma_0^2\tau^2/2 + \text{log}\left(1+\sum_{i=1}^{N}\frac{A_i}{A_0}e^{-\Delta m_i\tau+\Delta\Gamma_i^2\tau^2/2}\right),\\
    \approx &-m_0\tau+\Gamma_0^2\tau^2/2 + \sum_{i=1}^{N}\frac{A_i}{A_0}e^{-\Delta m_i\tau+\Delta\Gamma_i^2\tau^2/2},\\
    \intertext{using that $\log(1+x) \approx x$ for small $x$.}
    \frac{\partial\text{log}(G(\tau))}{\partial\tau} =& (-m_0+\Gamma^2_0\tau)+\sum_{i=1}^{N} \frac{A_i}{A_0}\left(-\Delta m_i + \Delta\Gamma_i^2\tau \right)e^{-\Delta m_i \tau + \Delta\Gamma_i^2\tau^2/2},\label{eq:fullmassfit}\\
    \frac{\partial^2\text{log}(G(\tau))}{\partial\tau^2} = &(\Gamma^2_0)+\sum_{i=1}^{N} \frac{A_i}{A_0}\left(\left(-\Delta m_i +\Delta\Gamma^2_i\tau\right)^2+\Delta\Gamma^2_i\right)e^{-\Delta m_i \tau + \Delta\Gamma_i^2\tau^2/2} \label{eq:fullwidthfit} .\\
    \intertext{In the following analysis we will be assuming only a single excited state contributes, i.e., $N=1$. Furthermore, when fitting the mass we take the narrow-width approximation $\Gamma^2\tau\ll m_0$, giving the fit functions}
    M(\tau) = & m_0+ Ae^{-\Delta m_i \tau }\label{eq:Mfit} ,\\
    \Gamma^2(\tau) = & \Gamma^2_0+ Ae^{-\Delta m_i \tau + \Delta\Gamma_i^2\tau^2/2}\label{eq:Gammafit} .
\end{align}
\section{Results}
\begin{figure}[t]
    \centering
    \begin{subfigure}{0.495\textwidth}
    \centering
    \includegraphics[width=\linewidth]{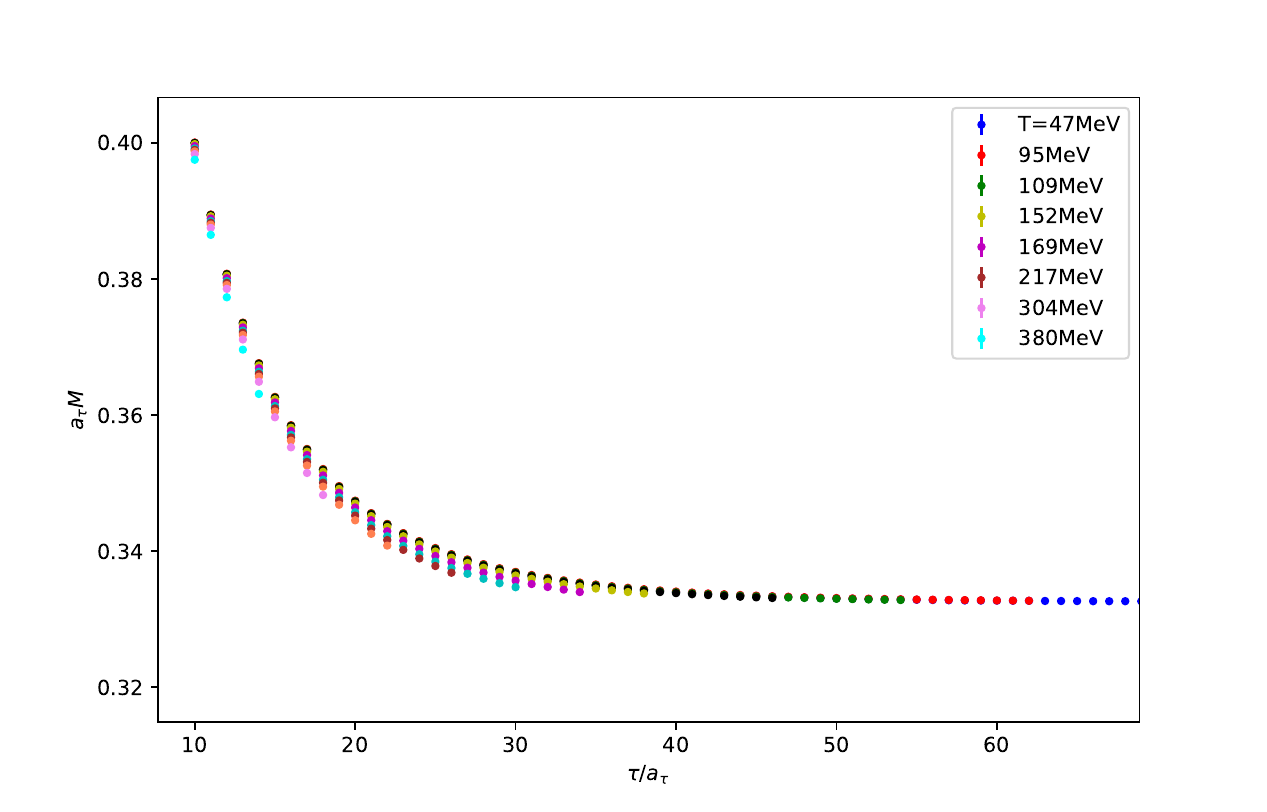}
    \end{subfigure}
    \centering
    \begin{subfigure}{0.485\textwidth}
    \centering
    \includegraphics[width=\textwidth]{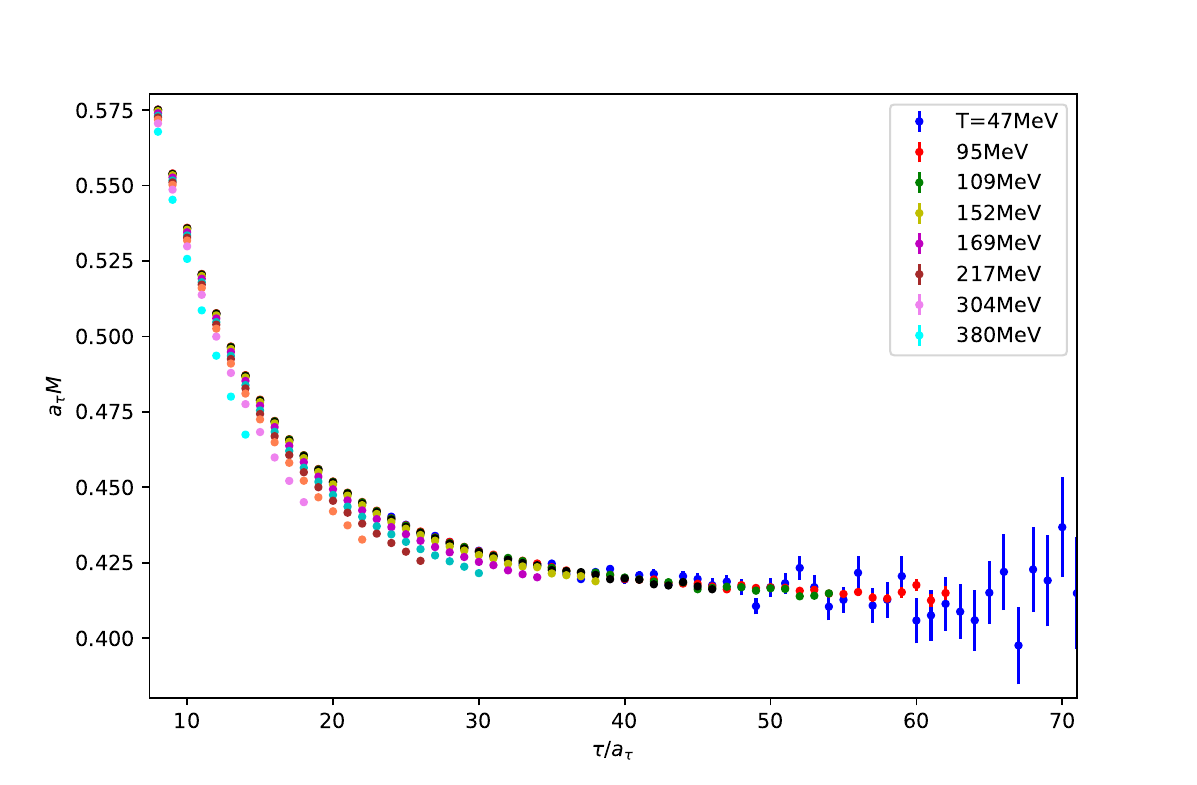}
    \end{subfigure}
    \caption{Left: $M_L(\tau)$ of the $\Upsilon$ at each temperature. Right: $M_L(\tau)$ of the $\cbo$ at each temperature.}
    \label{fig:mass_tau}
\end{figure}

\begin{figure}[t]
    \centering
    \begin{subfigure}{0.495\linewidth}
    \centering
     \includegraphics[width=\linewidth]{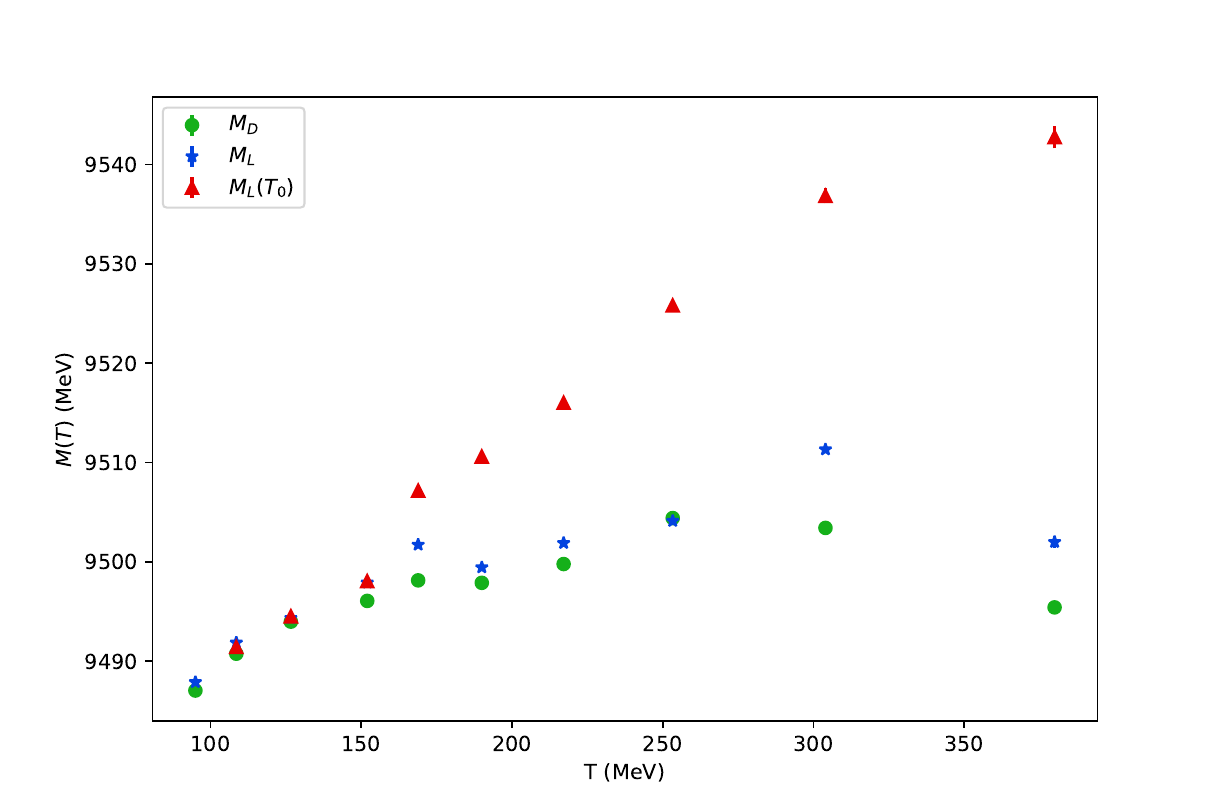}
    \end{subfigure}
    \centering
    \begin{subfigure}{0.485\linewidth}
    \centering
    \includegraphics[width=\linewidth]{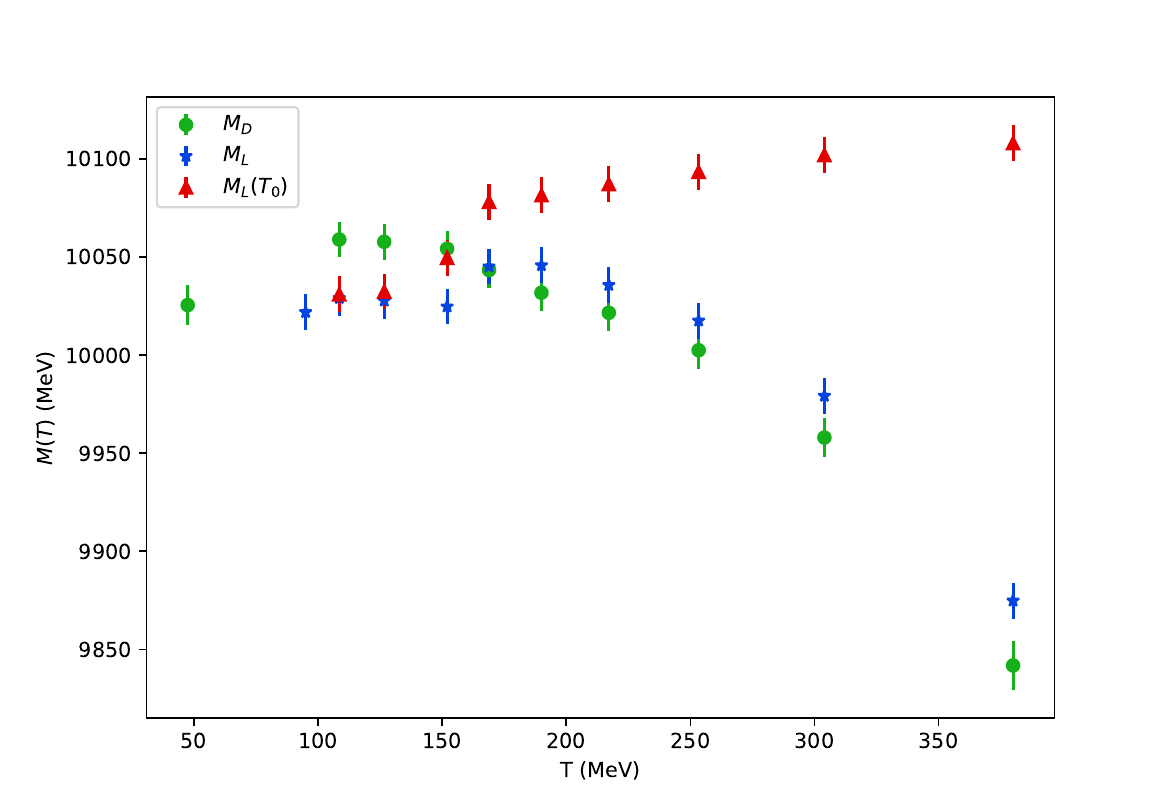}
    \end{subfigure}
    \caption{Left: $M_L$ (blue) from the fit for the $\Upsilon$ at each temperature and the corresponding $T_0$ (red) result. Right: $M_L$ (blue) from the fit for the $\cbo$ at each temperature and the corresponding $T_0$ (red) result. }
    \label{fig:mass_temp}
\end{figure}

In Figures \ref{fig:mass_tau} and \ref{fig:mass_temp} we show results for the bottomonium S-wave, $\Upsilon$, and the P-wave, $\cbo$, mass. We fit the mass to on exponential as in Equation \eqref{eq:Mfit}. A fourth order finite difference \cite{findiff}  was used for both the first and second order derivatives. $M(\tau)$ has the expected behaviour in that it is similar to that of the standard effective mass \cite{Aarts:2014cda,Offler:2019eij}. For the $\Upsilon$ we clearly see a plateau approaching a constant for temperatures below $150$MeV. For the higher temperatures above $150$MeV are approaching plateau but are still decreasing at the maximum value of $\tau$, Figure \ref{fig:mass_tau}, with a larger decrease for the higher temperatures indicating thermal effects. For the $\cbo$, $M(\tau)$ is quite noisy at large $\tau$, with some linear effects at large $\tau$ indicating the narrow width approximation may not be as suited to the $\cbo$ as it is with the $\Upsilon$. 

Figure \ref{fig:mass_temp} shows the mass as a function of temperature where we have included the additive NRQCD energy shift, which is $E_0 =7463$MeV for these ensembles, set by the $\Upsilon(1S)$ mass. The zero temperature results are in agreement with experiment:$M(\Upsilon)=9455(10)$MeV and $M(\cbo) = 9965(79)$MeV compared to the Particle Data Group values of 9460MeV and 9892MeV respectively \cite{ParticleDataGroup:2024cfk}. We see an increase in mass with increasing temperature for the $\Upsilon$ in agreement with some previous studies by FASTSUM \cite{Aarts:2011sm}. The apparent increase could be a kinematic effect due to the short temporal range at high temperatures \cite{Kelly:2018hsi, Kim:2018yhk}. This suggests that the ground state may not be dominating at late times, which would lead to an overestimate of the ground state mass. To disentange potential kinematic effects from genuine thermal effects we perform a zero temperature, $T_0$, analysis where we take the first $N_\tau$ points of the zero temperature correlator, $N_\tau$ corresponding to each temperature and repeat the analysis presented above. This analysis suggests that kinetic effects cause the increse in mass as the $T_0$ data does not experience any thermal effects. We also see a negative mass shift from the $T_0$ analysis, implying there is a decrease in the upsilon mass, as seen in \cite{Kim:2018yhk}. The $T_0$ results are shown in red in Figure \ref{fig:mass_temp}. The $\cbo$ is decreasing with increasing temperature as we can see from the fits and the $T_0$ analysis. 

\begin{figure}[t]
    \centering
    \begin{subfigure}{0.485\textwidth}
    \centering
    \includegraphics[width=\linewidth]{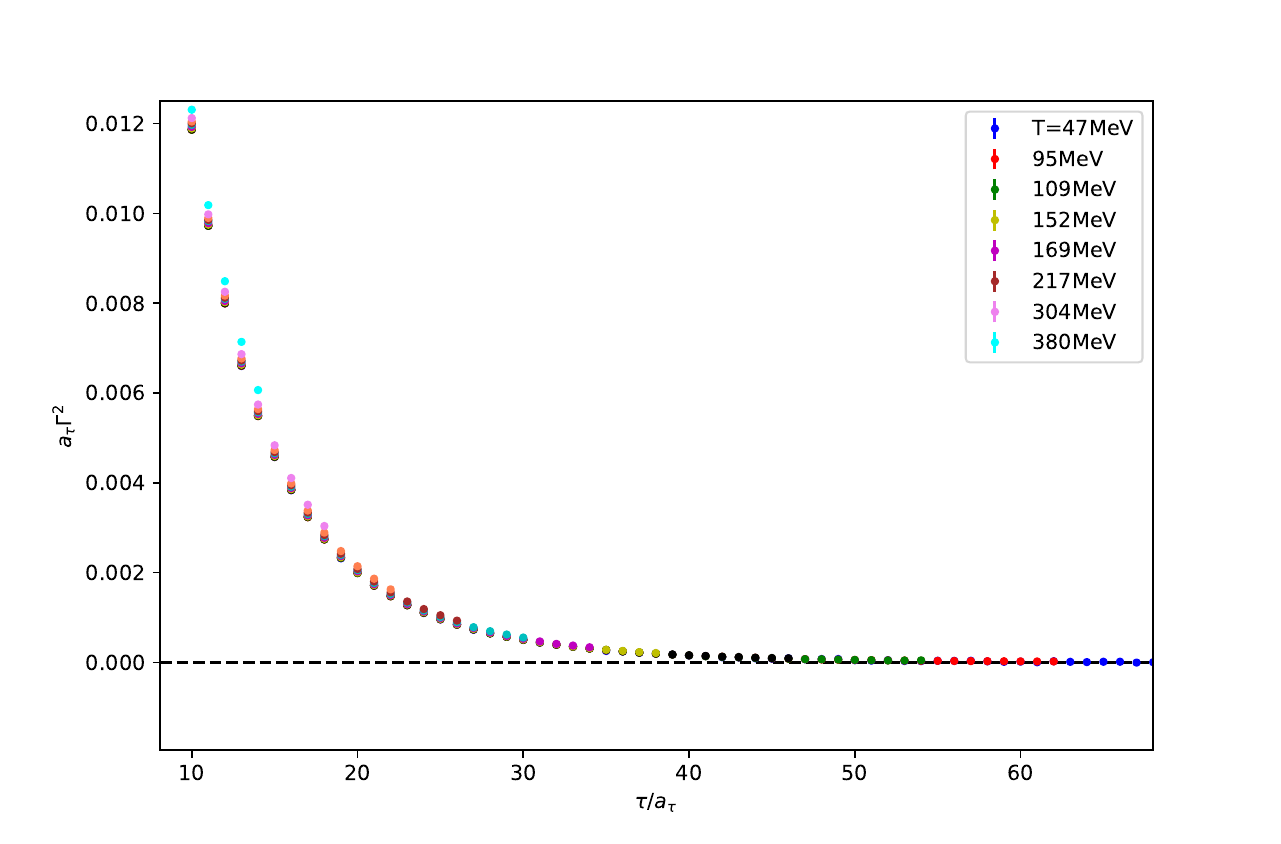}
    \end{subfigure}
    \centering
    \begin{subfigure}{0.495\textwidth}
    \centering
    \includegraphics[width=\textwidth]{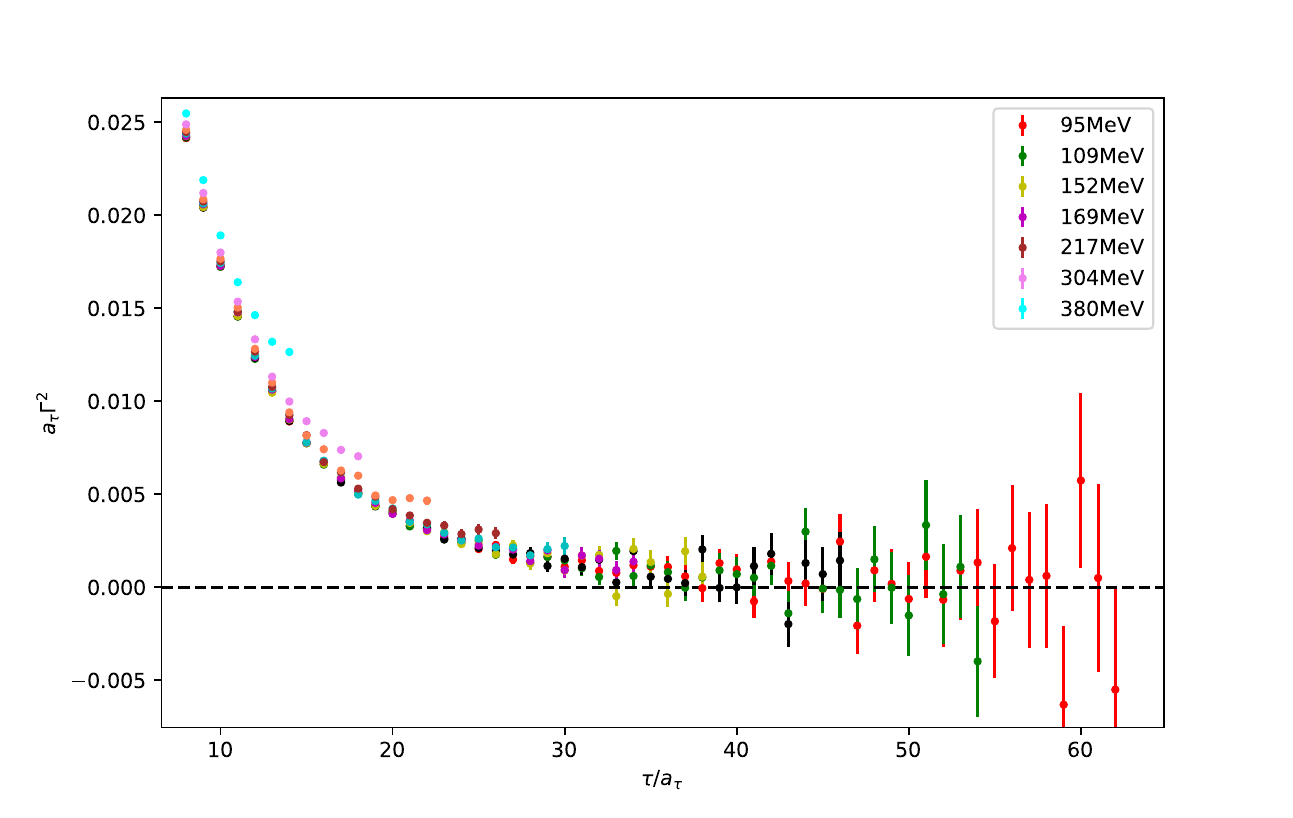}
    \end{subfigure}
    \caption{Left: $\Gamma^2_L(\tau)$ of the $\Upsilon$ at each temperature. Right: $\Gamma^2_L(\tau)$ of the $\cbo$ at each temperature. }
    \label{fig:width_tau}
\end{figure}

\begin{figure}
    \centering
    \begin{subfigure}{0.495\linewidth}
    \centering
     \includegraphics[width=\linewidth]{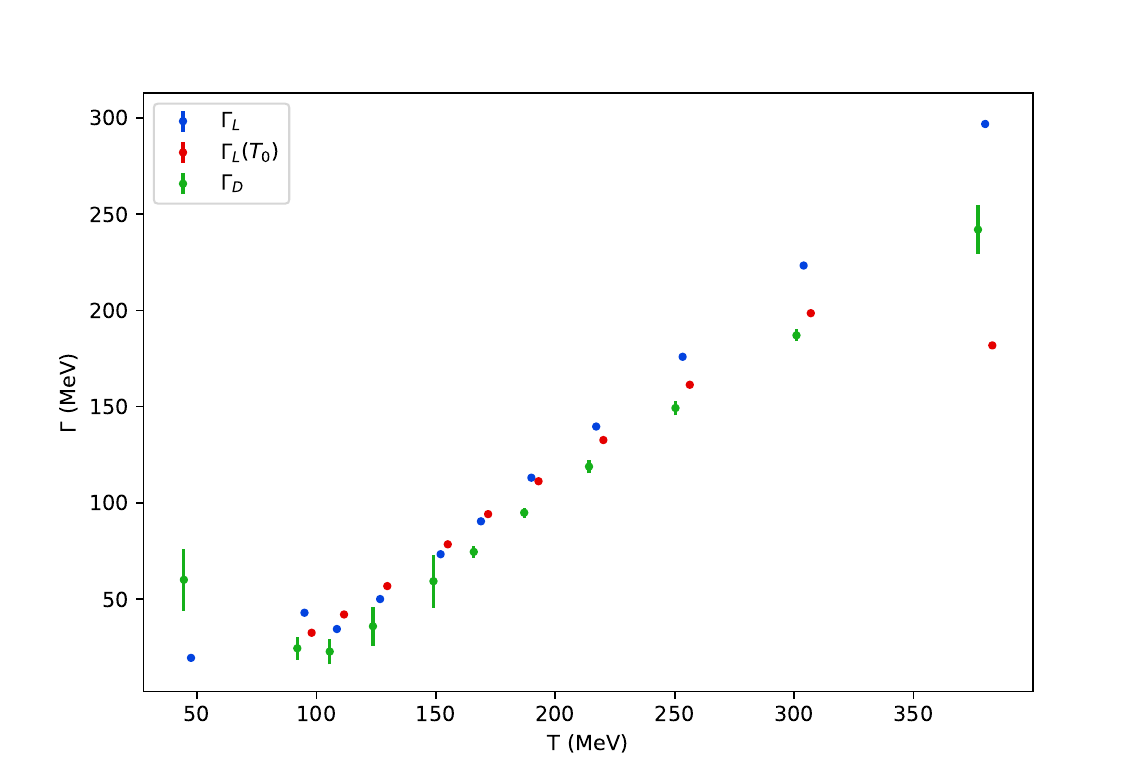}
    \end{subfigure}
    \centering
    \begin{subfigure}{0.495\linewidth}
    \centering
    \includegraphics[width=\linewidth]{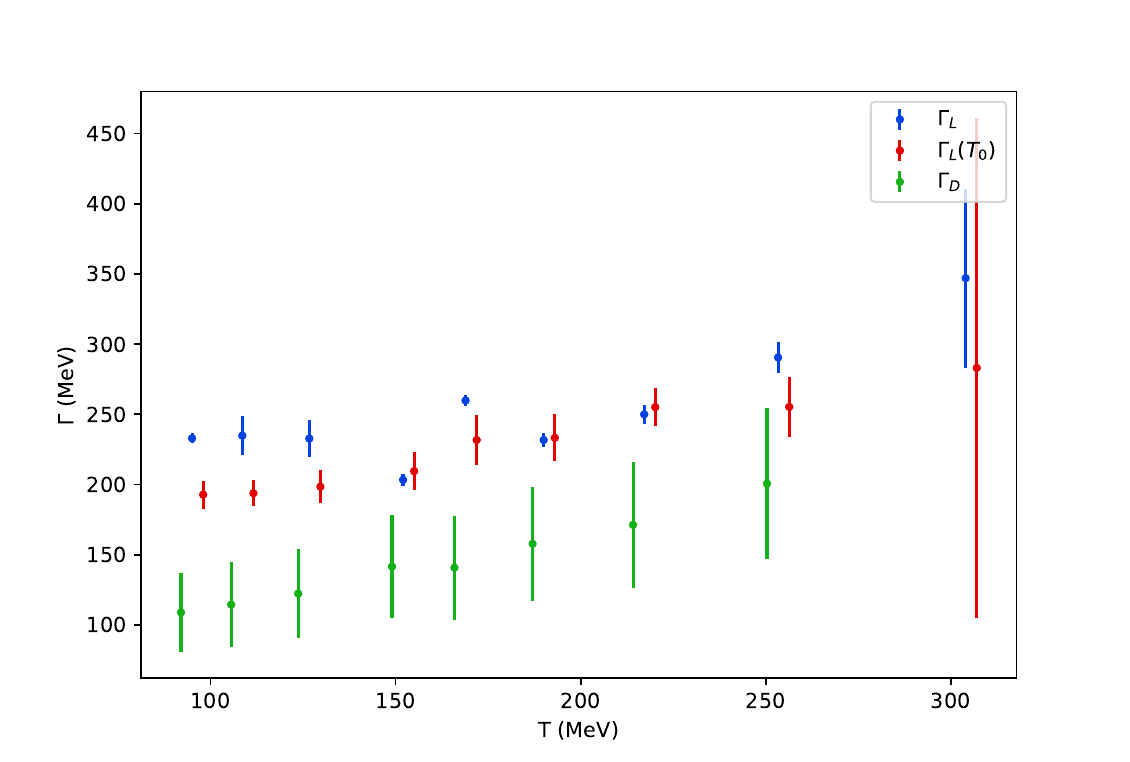}
    \end{subfigure}
    \caption{Left: $\Gamma_L$ (blue) and $\Gamma_D$ (green) from the fit for the $\Upsilon$ at each temperature and the corresponding $T_0$ (red) result. Right: $\Gamma_L$ (blue) from the fit for the $\cbo$ at each temperature and the corresponding $T_0$ (red) result. Points offset for clarity}
    \label{fig:width_temp}
\end{figure}

For the width we see in Figure \ref{fig:width_tau} that the $\Upsilon$ width is consistent with zero at zero temperature, has a plateau for the lower temperatures below $150$MeV and is still decreasing for the higher temperatures above $150$MeV indicating thermal effects. We see similar behaviour for the $\cbo$ although there is a lot of noise in the data at large $\tau$. We fit $\Gamma^2(\tau)$ for each temperature to Equation \eqref{eq:Gammafit}, shown in Figure \ref{fig:width_temp}. We see a clear increase in $\Gamma$ for both the $\Upsilon$ and $\cbo$ as the temperature increases. We also repeated the zero temperature the analysis on the width and found that we see an increase in thermal width with temperature. 

All errors were done using a bootstrap analysis. The systematic errors were calculated using Equation \eqref{eq:Gammasquared_weightedaverage} to find $\Gamma$ and Equation \eqref{eq:GprimeoverG} for the mass and \eqref{eq:Gammasquared_weightedaverage} to find $\Gamma$.
The systematic errors were calculated using Equation \eqref{eq:GprimeoverG} for the mass and \eqref{eq:Gammasquared_weightedaverage} for the width, as shown in Figures \ref{fig:mass_temp} and \ref{fig:width_temp} respectively.
The systematics are under control at low temperature for the $\Upsilon$, while the discrepancies for the $\cbo$ need to be considered further. 


\section{Conclusions}
We have presented a new method to finding the spectra of quarkonia that allows us to avoid the inverse problem. The mass results are in agreement with experiment at zero temperature. We see an increase of $\Gamma$ with temperature and a decrease in $M$ with temperature. Without having done a zero temperature analysis the data agrees with previous FASTSUM results that found a increase in mass \cite{Aarts:2011sm, Aarts:2014cda}. However the zero temperature analysis suggests that the apparent increase is likely due to the limited number of data points at higher temperature since it also sees the increase which can't be attributed to thermal effects.  

An outstanding question is if we should be taking the limit as $\Gamma^2 \to 0$ for the $\cbo$, or whether the width is too large for this assumption to hold. For further study, we will consider fitting the mass to Equation \eqref{eq:fullmassfit} and the width to Equation \eqref{eq:fullwidthfit} for $N\neq 1$. A multi-exponential fit analysis has been done in \cite{RyanProceedings}. We will be repeating these results on new ensembles with twice the number of temporal lattice sites for each temperature as this study. This will allow us to compare the effect the number of temporal sites has on the method. 
\section*{Acknowledgements}
RHD has been supported by Taighde Éireann – Research Ireland under Grant number  

GOIPG/2024/3507. This work used the DiRAC Extreme Scaling service at the University of Edinburgh, operated by the Edinburgh Parallel Computing Centre on behalf of the STFC DiRAC HPC Facility (www.dirac.ac.uk). This equipment was funded by BEIS capital funding via STFC capital grant ST/R00238X/1 and STFC DiRAC Operations grant ST/R001006/1. DiRAC is part of the National e-Infrastructure. This work was performed using PRACE resources at Cineca via grants 2015133079 and 2018194714. We acknowledge the support of the Supercomputing Wales project, which is part-funded by the European Regional Development Fund (ERDF) via Welsh Government,
and the University of Southern Denmark for use of computing facilities. 

\bibliographystyle{JHEP}
\bibliography{lattice24}

\end{document}